# Forecasting carbon-risk hotspots ahead of the extreme 2026 El Niño

Xiaomeng Du[1], Runda Jiang[1], Philippe Ciais[2], Pierre Friedlingstein[3,4], Stephen Sitch[3], Mike O'Sullivan[3], Jefferson Goncalves de Souza[3], Tim Stripp[3], Eloi Lindas[2], Piyu Ke[2], Yi Xi[2], Minxuan Sun[1], Xiaoran Wu[1], Mengjie Han[1], Shanshan Song[1], Qiang Zhang[1] and Wei Li[1]*

[1]Department of Earth System Science, Ministry of Education Key Laboratory for Earth System Modeling, Institute for Global Change Studies, Tsinghua University, Beijing 100084, China

[2]Laboratoire des Sciences du Climat et de l'Environnement, LSCE/IPSL, CEA-CNRS-UVSQ, Université Paris-Saclay, 91191 Gif-sur-Yvette, France

[3]Faculty of Environment, Science and Economy, University of Exeter, Exeter, EX4 4QF, UK

[4]Laboratoire de Météorologie Dynamique, Institut Pierre-Simon Laplace, CNRS, École Normale Supérieure, Université PSL, Sorbonne Université, École Polytechnique, Paris, France

*Correspondence to Wei Li: wli2019@tsinghua.edu.cn

## Abstract

A potentially super El Niño is forecasted in 2026, raising the need to anticipate land carbon-cycle risks before its impacts unfold. We develop a prospective early-warning framework that uses dynamic global vegetation models driven by an ensemble of ECMWF seasonal climate forecasts and machine-learning burned area predictions. Forecasts indicate a 0.23 ± 0.21 PgC reduction in global net biome productivity in the second half of 2026, with hotspots of sink weakening in eastern South America and elevated fire risk in Southeast Asia.

## Main

An El Niño event is emerging since June 2026 and is projected to intensify further through the second half of the year, potentially approaching the intensity comparable to the strongest events in recent decades[1]. Multiple operational forecast systems, such as NOAA forecasts and independent multi-model ensembles, converge on the fact that this event will reach the very strong category and peak during October-December 2026[2]. Anticipating its ecosystem consequences before they fully develop is therefore important for timely and targeted risk management.

El Niño events substantially disrupt the global carbon cycle through widespread tropical drought and heat anomalies and increased fire activity, both of which lead to reduction in the global land carbon sink[3,4]. During the previous 1997-1998, 2015-2016 and 2023-2024 El Niño events, terrestrial carbon sinks weakened by about 50%, driven by reduced vegetation productivity and increased ecosystem carbon losses[5,6,7]. Yet, terrestrial carbon-cycle responses to extreme El Niño events vary strongly across regions, owing to differences in the magnitude, timing, and persistence of climate anomalies and ecosystem sensitivity[8]. While seasonal forecasts have increasingly enabled early warnings of climate-related hazards, such as droughts and floods, their implications for the terrestrial carbon balance have not been used yet. Combining dynamic global vegetation models (DGVMs) with seasonal climate forecasts provides a means to translate climate outlooks into prospective estimates of carbon-cycle risk. Here, we use ECMWF seasonal climate forecasts to drive two DGVMs (ORCHIDEE-MICT[9] and JULES[10,11]), incorporating machine-learning-based burned area predictions to anticipate changes in the terrestrial carbon balance and fire responses to the developing 2026 El Niño. We used ECMWF ensemble forecasts initialized on 1 June 2026 as climate forcing for the second half of 2026. The 51-member ensemble forecasts extend seven months ahead and represent plausible climate trajectories[12]. Building on continuous historical simulations from 1960 to 2025 using the ERA5 forcing[13], we run simulations with the ERA5 forcing[14,15] for January–June 2026 and switch to each of the 51 ECMWF ensemble members[12] for July–December 2026. Our forecast simulations identify regions likely to experience enhanced carbon losses and fire activity, offering an opportunity to anticipate disruptions to the land carbon cycle before their full impacts unfold.

By July 2026, multiple ENSO indices (e.g., the Oceanic Niño Index and Multivariate ENSO Index) indicate that the developing El Niño was already comparable in intensity to, or stronger than, the major 1997–1998 and 2015–2016 events at the same stage of development (Fig. S1). This extreme event is already accompanied by widespread land climate anomalies (climate and carbon flux anomalies are calculated relative to the corresponding monthly means during 2016–2025 reference period, excluding El Niño years). During April–June 2026, global land temperature anomalies exceeded those of previous extreme El Niño events, peaking at approximately +0.39 °C in June and surpassing the 90th percentile for the corresponding month during the reference period. (Fig. 1a, Fig. S1). For July–December 2026, the mean of the 51 ECMWF ensemble members projects near-neutral global land temperature anomalies, similar to those in 2015. However, substantial ensemble spread leaves open the possibility of November anomalies approaching +0.89°C, comparable to those in 2023 (Fig. 1a). Tropical warming is more consistent across the ensemble, with 48 of the 51 members projecting peak tropical-mean temperature anomalies above +0.3°C during July–December and the ensemble median peak anomaly reaching +0.59°C (Fig. S2). The largest positive temperature anomalies are projected over tropical America (Fig. S3).

Seasonal ECMWF forecasts project pronounced global precipitation deficits, with the strongest negative anomalies in August–September (Fig. 1b). All 51 ensemble members predict global precipitation anomalies below the reference-period 10th percentile (Fig. 1b and S4). These deficits are driven primarily by the tropics, where precipitation during July–December 2026 is projected to be 11.7 ± 2.3% lower than the reference-period mean level (Fig. S4). Below-average precipitation is projected across 50.9% of tropical land, with the largest negative anomalies concentrated over tropical America and Africa (Fig. S5). Together with the projected tropical warming, these deficits suggest concurrent warm and dry conditions in the tropics during the peak phase of the 2026 El Niño.

Relatively cool and wet conditions supported widespread positive NBP anomalies (enhanced sink) during the first half of 2026, but this is projected to reverse during the second half of the year (Fig. 1c). The global NBP anomaly in 2026 followed a similar trajectory to those in 2015 and 2023, becoming progressively more negative through the second half of the year and reaching its most negative anomalies in November (−0.12 ± 0.08 PgC $mon^{-1}$) and December (−0.13 ± 0.03 PgC $mon^{-1}$). The cumulative global land NBP anomaly for July–December 2026 is projected to reach −0.23 ± 0.21 PgC, compared with −0.95 ± 0.03 and −1.39 ± 0.02 PgC over the same period in 2015 and 2023, respectively. Although the ensemble-mean anomaly is less negative than 2015 and 2023, the most negative anomaly among the 51 members (−1.0 ± 0.3 PgC) is comparable to the 2015 value, suggesting a potentially similar weakening of the land carbon sink (Fig. 1d and Fig. S6). The regional contributions also differ from those in previous El Niño years. Whereas the tropics dominated the negative global NBP anomalies in 2015 and 2023, the projections suggest a shift towards comparable contributions from the tropics and the northern extra tropics in 2026 (Fig. 1f).

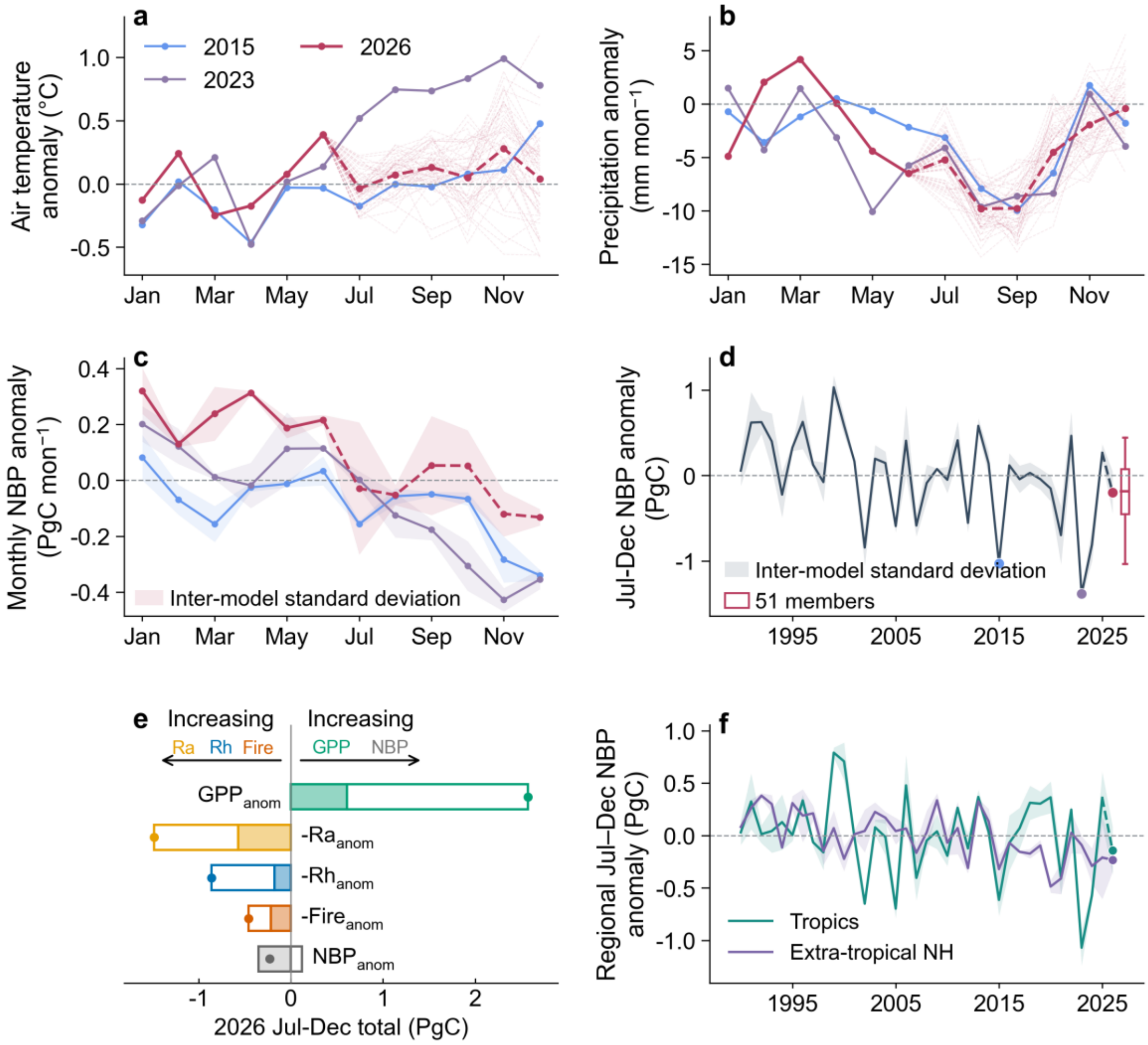

**Figure 1. Climate anomalies and terrestrial carbon cycle responses to 2026 El Niño event.** (a–b) Monthly anomalies in global mean land temperature and precipitation. Anomalies for 2015, 2023 and January–June 2026 are based on ERA5[14,15] while those for July–December 2026 are derived from ECMWF seasonal forecasts[12]. The pale red dashed lines show individual forecasts from the 51 ECMWF ensemble members. (c) Monthly anomalies in global total net biome productivity (NBP), with shading indicating ±1 standard deviation across dynamic global vegetation models (DGVMs). Negative NBP anomalies indicate a weaker carbon sink or a stronger carbon source relative to the reference period (2016-2025, excluding El Niño years). (d) Detrended cumulative global NBP anomalies for July–December of each year from 1990 to 2026. Dots highlight the extreme El Niño years 2015, 2023 and 2026. The shading indicates ±1 standard deviation across DGVMs. The boxplot for 2026 represents the distribution of July–December NBP anomalies across the 51 ECMWF ensemble members. (e) Contributions of carbon flux components to the cumulative global NBP anomaly during July–December 2026. Stacked bars show anomalies in gross primary productivity (GPP), autotrophic respiration (Ra), heterotrophic respiration (Rh) and fire emissions (Fire). Note that respiration and fire emission anomalies are shown with reversed signs to reflect their contributions to the NBP anomaly ($NBP_{anom} = GPP_{anom} - Ra_{anom} - Rh_{anom} - Fire_{anom}$). Dots represent global totals, while filled bars represent tropical contributions and open bars represent contributions from regions outside the tropics. (f) Detrended cumulative NBP anomalies for July–December of each year from 1990 to 2026 in the tropics (23.5°S–23.5°N; green) and extra-tropical Northern Hemisphere (23.5°–90°N; purple). Shading indicates ±1 standard deviation across DGVMs.

To identify the processes driving the projected NBP anomalies, we further decompose the NBP anomalies into anomalies of gross primary production (GPP), autotrophic respiration (Ra), heterotrophic respiration (Rh) and fire emissions. At the global scale, enhanced ecosystem respiration was the dominant contributor to the projected NBP reduction during the second half of 2026, partly offset by increased GPP in temperate and boreal regions where warmer conditions favored photosynthetic uptake (Fig. 1e). In the tropics, the increase in carbon release (respiration and fire emissions) exceeded the increase in carbon uptake, resulting in an overall negative NBP anomaly. Despite a positive GPP anomaly in the tropics, the tropical forests experiencing hotter and drier conditions showed declining GPP associated with suppressed photosynthetic activity (Fig. S7). Fire emissions provide an additional contribution to negative NBP anomalies in the tropics. These results indicate that the projected carbon sink weakening during the 2026 El Niño arises from the enhanced carbon release through ecosystem respiration and fire emissions (Fig. 1f). The NBP reduction of 0.23 ± 0.21 PgC is due to an increase of 1.48 ± 1.61, 0.86 ± 1.30 and 0.46 ± 0.02 PgC in Ra, Rh and fire emissions respectively, with the GPP increase of 2.57 ± 2.72 partly offsetting that reduction.

To assess land carbon-cycle responses under forecast conditions closest to observations, we selected the forecast members with the lowest mean absolute error (MAE) in temperature and precipitation, respectively, using ERA5 data for July–August 2026 as the reference (Fig. S8-9). The member with global temperature closest to ERA5 projected a cumulative NBP anomaly of −0.35 ± 0.37 PgC for the second half of 2026, more negative than the ensemble mean (Fig. S10). The difference was larger in the tropics, where the member that most closely matched temperature projected a cumulative anomaly of −1.10 ± 0.27 PgC, i.e., 0.75 PgC below the ensemble mean (Fig. S11). Notably, ERA5 temperatures during July–August exceeded those of all forecast members, and precipitation fell below that of the driest forecast members in the tropics (Fig. S9). This suggests that the actual weakening of the terrestrial carbon sink in the second half of 2026 could be more negative than that projected under the most extreme ECMWF conditions.

The 2026 super El Niño is projected to pose substantially greater risks of localized carbon sink weakening than the 2015 and 2023 El Niño events, despite weaker net negative anomalies globally and across the tropics owing to spatial compensation. Approximately 11.4% of global land area is projected to experience more negative NBP anomalies than in both 2015 and 2023, while 19.6% is projected to exceed at least one of them. These proportions are higher in the tropics, reaching 16.4% and 26.1%, respectively (Fig. 2a, b). Tropical South America makes the largest contribution to the projected weakening of the global terrestrial carbon sink during July–December 2026, with an NBP reduction of 0.24 ± 0.1 PgC, accounting for approximately 44% of the global decline (Fig. 2c). Negative NBP anomalies are projected by 44 of the 51 ensemble members (86.3%) in this region, indicating strong agreement on the direction of change. Within tropical South America, the largest negative anomalies are concentrated in Brazil, extending across humid tropical forests in northeastern Brazil and grassland and savanna ecosystems in central Brazil (Fig. 2c). NBP reductions are also projected for southern Africa, with affected areas including grassland and savanna ecosystems (Fig. 2c).

The forecasts identify region-specific windows of elevated carbon-loss risk during the developing 2026 El Niño. Eastern Brazil faces the highest risk during September–December (Fig. S12). Other regions also show distinct windows of projected carbon sink weakening, including central North America during July–August, southern Africa during September–November, and eastern Australia during November–December (Fig. S12). These regionally resolved forecasts provide early-warning information on where and when terrestrial carbon sink may weaken, enabling proactive monitoring and management to mitigate carbon losses and strengthen ecosystem resilience as the 2026 El Niño evolves.

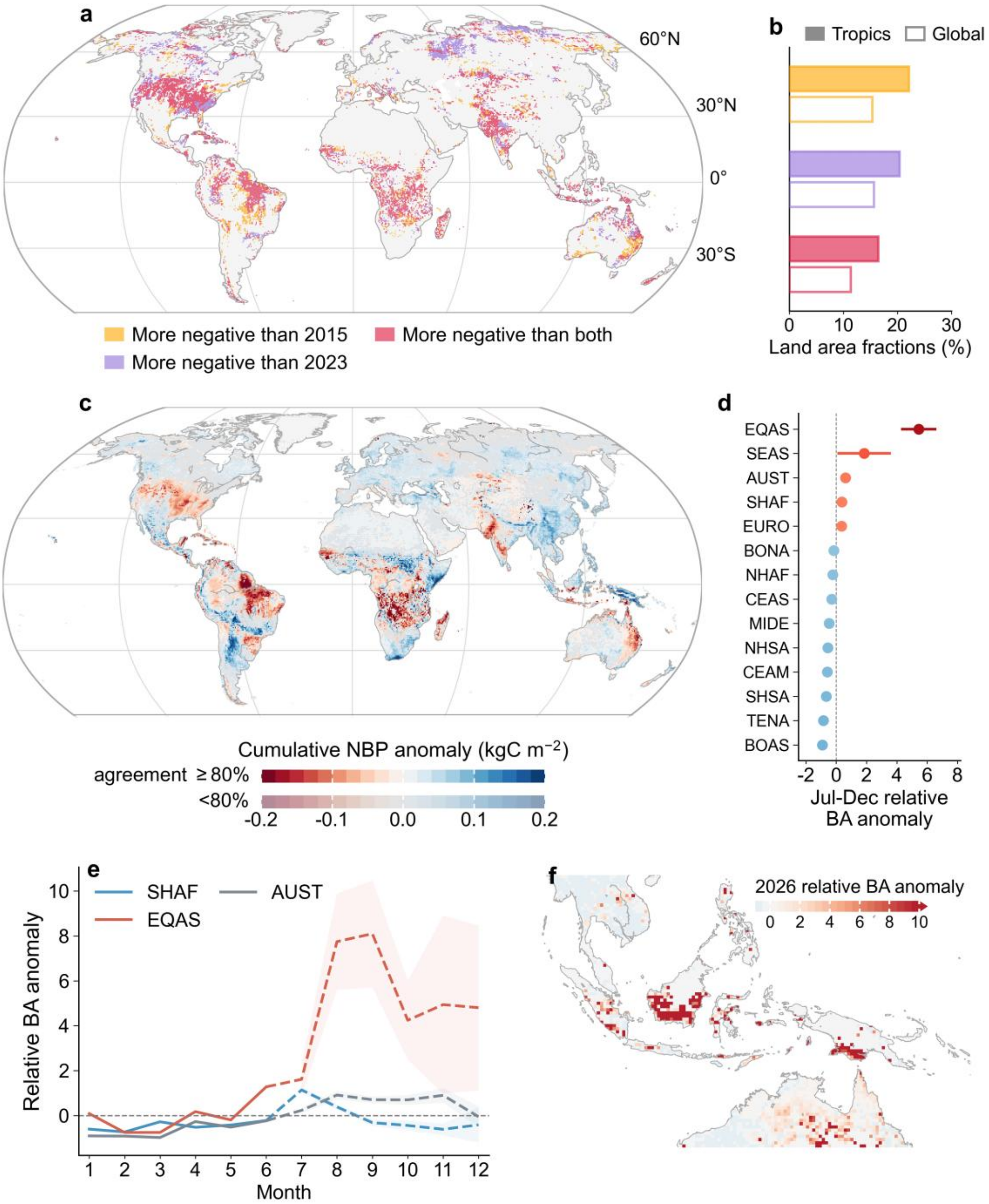

**Figure 2. Spatial patterns of projected net biome production and burned area anomalies during the second half of 2026.** (a-b) Spatial distribution (a) and land-area fractions (b) of NBP anomalies in the second half of 2026 that are more negative than those in the corresponding periods of 2015 and 2023. Only grid cells where the 2026 NBP anomaly is negative by more than 5% are included. (c) Spatial distribution of cumulative net biome production (NBP) anomalies from July to December 2026. Dark colors indicate grid cells where at least 80% of the 51 ECMWF ensemble members agree on the sign of the anomaly. Areas with lower ensemble agreement (less than 80%) are overlaid in gray. (d) Cumulative relative burned area (BA) anomalies from July to December 2026 across the Global Fire Emissions Database (GFED) regions. The relative BA anomaly is calculated as the ratio of 2026 BA anomaly to the mean BA during the 2016-2025 reference period, excluding El Niño years (i.e., $\text{relative BA anomaly} = (BA_{2026} - \overline{BA_{ref}}) \,/\, \overline{BA_{ref}}$). Error bars represent the standard deviation across the machine-learning predictions generated by the 51 ECMWF ensemble members. GFED regions are shown in Fig. S18 as follows: BONA (boreal North America), EURO (Europe), MIDE (Middle East), BOAS (boreal Asia), CEAS

(Central Asia), SEAS (Southeast Asia), EQAS (equatorial Asia), AUST (Australia), SHAF (Southern Hemisphere Africa), NHAF (Northern Hemisphere Africa), SHSA (Southern Hemisphere South America), NHSA (Northern Hemisphere South America), CEAM (Central America), TENA (temperate North America). (e) Relative burned area anomalies during 2026 in tropical regions with high fire risk. Shading denotes the standard deviation across the machine-learning predictions generated by the 51 ECMWF ensemble members. (f) Spatial distribution of relative burned area anomalies in equatorial Asia during 2026 relative to the reference period mean. Red indicates areas where projected BA exceeds the reference-period mean level.

As El Niño events are typically associated with elevated fire activity, we further forecasted regional burned area during the 2026 El Niño based on a machine-learning framework[16]. Global burned area is projected to increase substantially during July–August 2026, peaking in July when burned area exceeds the 90th percentile of the reference period (Fig. S13). This early-season increase is primarily driven by enhanced fire activity in Southern Africa (SHAF), where the burned area in July is projected to be approximately twice the 2016-2025 (excluding El Niño years) average (Fig. S14).

Although global burned area anomalies are projected to return to near-neutral levels after September, substantial regional fire risks persist during the second half of 2026 (Fig. 2d and S13-14). Equatorial Asia (EQAS) exhibits the highest relative burned area anomaly (i.e., the ratio of 2026 BA anomaly to the reference period mean level), with cumulative July–December burned area being 5.6 ± 1.2 times higher than the 2016–2025 average (Fig. 2e). The burned area increases are concentrated in Indonesia, Papua New Guinea, and Malaysia, where high burned area anomalies largely overlap with peatland regions (Fig.2f, S15-16). Approximately 12.18 Mha of tropical forested peatlands are expected to experience elevated fire risk during the second half of 2026, highlighting the need for strengthened fire-risk management in Equatorial Asia peatlands, including maintaining hydrological conditions, preventing drainage-induced drying, and reducing human-driven ignition. Fire activity is also projected to increase in Australia (AUST), with burned area during the second half of 2026 approximately 61.8 ± 11.2% higher than the reference period mean level and primarily concentrated in northern Australian grassland and savanna ecosystems (Fig. 2e and S17). GFED5 NRT estimates provide independent support for the predicted burned area anomalies in Southeast Asia and Australia, with particularly pronounced agreement over peatland regions in Southeast Asia (Fig. S19-21).

The projected increase in tropical fire activity is associated with climate-driven changes in fire-conducive conditions, with enhanced vapor pressure deficit (VPD) emerging as the dominant factor in Equatorial Asia and fire weather playing a major role in Australia (Fig. S22, 23). These results highlight the tropics, particularly Equatorial Asia, as priority region for fire-risk monitoring during the 2026 El Niño event and underscore the importance of region-specific fire-risk assessment and management in extreme climate events.

Extreme climate events affect terrestrial carbon cycling through regionally distinct pathways shaped by ecosystem characteristics and prior climate stress. In tropical South America, drought-induced hydraulic damage, canopy loss and delayed mortality may persistently limit ecosystem productivity, reducing resistance to subsequent climate anomalies and thereby exacerbating carbon sink declines[17,18]. By contrast, carbon losses in Southeast Asia are strongly associated with higher fire risks. Extensive peatlands are highly vulnerable to drought-induced fires, with long-term land-use changes further degrading hydrological conditions and increasing fire susceptibility in this region[6,19]. Because fire emissions increase exponentially as water deficits accumulate[20,21], sustained monitoring of peatlands and forests in Southeast Asia should be prioritized for fire-risk management during El Niño events.

The uncertainty of ECMWF seasonal forecasts increased with lead time and differed substantially among variables (Supplementary Text S1, Fig. S24-29). Precipitation exhibited larger ensembles spread, particularly over tropical regions (Fig. S24). In contrast, temperature forecasts showed relatively smaller ensemble spread, with uncertainties mainly occurring over Northern Hemisphere mid-to-high latitudes (Fig. S24). Although precipitation uncertainty increased toward longer lead times, all 51 ensemble members consistently projected large-scale drying tendency during the second half of 2026 (Fig. S25-27), suggesting that the uncertainty primarily affects the magnitude and spatial details rather than the direction of the predicted climate anomalies. By integrating seasonal climate forecasts with process-based DGVMs and machine-learning fire predictions, we identify tropical America, particularly Brazil, as a hotspot of carbon sink weakening during the 2026 El Niño, and equatorial Asia, particularly Indonesia and Papua New Guinea, as a hotspot of increased fire activity. More broadly, this study provides a pathway from retrospective assessments toward anticipatory carbon-cycle forecasting, facilitating adaptive management under increasing climate extremes.

## Methods

### Climate forcing

We used ERA5 climate forcing[14,15] from January to June 2026 and ECMWF seasonal forecasts[12] initialized on 1 June 2026 for July–December 2026. The ECMWF forecasts provided near-surface air temperature, surface pressure, wind speed, precipitation, and downward shortwave and longwave radiation (Table S1). The ECMWF forecasts ensemble consists of 51 members, with each member representing a plausible evolution of the 2026 climate conditions. Since ERA5 reanalysis (~ June 2026) and ECMWF seasonal forecasts (June 2026 ~) are generated from different assimilation and prediction systems[12,15], their combination may introduce potential discontinuities during the transition from June to July 2026. We therefore evaluated the consistency between the two datasets during their overlapping June 2026 period and assessed seasonal forecast performance by comparing ECMWF hindcasts for 1993–2016 with ERA5 reanalysis (Supplementary Text S1, Fig. S26-29), which showed good agreement in the examined monthly mean variables. The ECMWF seasonal forecasts were regridded from the original 1° × 1° to 0.5° × 0.5° to match the resolution of the DGVM forcing. Temperature and precipitation anomalies were calculated relative to the mean value for the corresponding month during the 2016–2025 reference period, excluding El Niño years.

### Simulation setup

We projected terrestrial carbon-cycle responses to the 2026 El Niño by driving two dynamic global vegetation models (DGVMs), ORCHIDEE-MICT[9] and JULES[10,11], with all 51 members of the ECMWF seasonal forecast ensemble, yielding 102 simulations (2 DGVMs × 51 climate forecast members). For 2026, simulations used ERA5 forcing for January–June and branched into 51 ensemble members for July–December, capturing the spread associated with seasonal climate forecasts within each DGVM. Simulations were initialized from a continuous historical run forced with ERA5 from 1960 to 2025[13,22], with atmospheric $CO_2$ concentrations from NOAA used in the Global Carbon Budget assessments[23] and land use prescribed from LUHv2h[24] and fixed at the 2010 level. Land-use change was therefore not explicitly represented in these simulations. The simulations were replicated across the 51 ECMWF ensemble members for July–December 2026. Each member was forced with its corresponding ECMWF forcing and the associated machine learning-predicted burned area driven by this ECMWF forcing, while atmospheric $CO_2$, land use, model parameters, and other non-meteorological forcings were kept identical across members. Therefore, differences among ensemble members reflect the impacts of different seasonal meteorological trajectories and their associated fire activity.

We analyzed anomalies in NBP, gross primary productivity (GPP), autotrophic respiration (Ra), heterotrophic respiration (Rh), and fire emissions during the 2026 El Niño event. Anomalies were calculated by subtracting the monthly mean values over the 2016–2025 reference period (excluding El Niño years) from each corresponding month in 2026. Positive NBP anomalies indicate a stronger carbon sink or a weaker carbon source relative to the reference period. We evaluated the DGVMs' ability to reproduce terrestrial carbon sink anomalies against atmospheric inversion estimates during the strong El Niño events of 1997–1998 and 2015–2016, and the DGVMs captured the interannual variability in the terrestrial carbon sink well (Supplementary Text S2, Figs. S30 and S31).

**Burned area prediction**

Here we developed a machine-learning (ML) framework[16] to predict monthly burned area at 0.5° × 0.5° resolution for 2021–2026 (Fig. S32). Burned areas were predicted at the monthly scale separately on forest and non-forest areas within each of the 14 Global Fire Emission Dataset (GFED)[25,26] regions (Fig. S18). We first used a random forest classifier (RFC) to predict the occurrence of fire and then applied a long short-term memory (LSTM) regressor to estimate burned area fraction for grid cells with fire occurrence. The burned area observation used as training targets was derived from the FireCCI51 burned-area product (2003–2020)[27,28]. Burned pixels overlapping cropland classes were excluded before model training using the CCI land-cover layer provided with FireCCI51[28]. The explanatory variables for ML model training and prediction were selected to represent their contributions to burned area, including climatic conditions, vegetation status, and human activities (Table S2). Climatic predictors including daily maximum and minimum temperature, precipitation, wind speed, vapor pressure deficit (VPD), and the Fire Weather Index (FWI), were extracted or calculated at the daily scale and averaged to the monthly scale. Human-related predictors included population density and land-use fractions, and leaf area index (LAI) was used to represent vegetation status.

For each GFED region and burned area on forest/non-forest respectively, the random forest classifier was first applied to determine whether fire would occur in each grid cell, and the long short-term memory regressor was then used to predict the burned area fraction in grid cells with fire occurrence. ML models were evaluated using five expanding forward-validation folds, with 2013–2017 as the validation years and each model trained only on data preceding the corresponding validation year. During model development, we tuned "max_depth" and "n_estimators" for the random forest classifiers, evaluating candidate models based on classification accuracy, the precision–recall balance for the fire class, F1 score, and temporal stability. For the LSTM regressors, we tuned "hidden_sizes", "learning_rate", and "epochs", with hyperparameters selection based on $R^2$. After selecting the optimal hyperparameters, the final models were then retrained using all data from 2004 to 2020 and used to predict burned area for 2021–2026. Model performance was evaluated through leave-one-year-out validation and comparison with independent satellite products[27,28,29,30] (Supplementary Text S3, Fig. S33-38).

Monthly relative burned area anomalies for 2026 were calculated relative to the corresponding monthly mean for reference period (2016–2025, excluding El Niño years), using FireCCI51 observations for 2016–2020 and machine-learning-based burned area predictions for 2021–2025. The relative burned-area anomalies were calculated as:

$$\mathit{Relative}\ \mathrm{BA}_{\mathrm{anomaly}} = \frac{\mathrm{BA}_{2026} - \mathrm{BA}_{\mathrm{ref}}}{\mathrm{BA}_{\mathrm{ref}}}$$

where $\mathrm{BA}_{\mathrm{ref}}$ is the mean burned area for the corresponding month during the 2016–2025 reference period, excluding El Niño years. To assess peatland fire risk under El Niño, we resampled the peatland data[13] to a 0.5° grid and estimated peatland burned area for each grid cell as burned-area fraction × peatland fraction × grid-cell area. We also independently estimated the 2026 burned-area anomalies using the Global Fire Emissions Database 5 near-real-time extension (GFED5 NRT)[31,32,33] and compared them with our machine-learning predictions.

## Data availability

El Niño indices are available from NOAA at https://psl.noaa.gov/enso/dashboard.html. ECMWF seasonal ensemble forecasts are available from the Copernicus Climate Data Store at https://cds.climate.copernicus.eu/datasets/seasonal-original-single-levels. Other source data are available at https://doi.org/10.6084/m9.figshare.34063923.

## Code availability

The scripts used to generate the results were developed using Python 3.8.19 and are available on Figshare via https://doi.org/10.6084/m9.figshare.34063923.